\documentclass[sigconf]{acmart}

\usepackage{graphicx}  
\usepackage{soul}
\usepackage{amsmath}
\usepackage{booktabs}
\usepackage{tablefootnote}
\usepackage{algorithm}
\usepackage{algorithmic}
\usepackage{afterpage}
\usepackage{tabularx}
\usepackage{atbegshi}
\usepackage{dblfloatfix}
\usepackage[utf8]{inputenc}
\usepackage{fullpage}
\usepackage{epsfig, endnotes, listings, graphicx, subcaption}
\usepackage{url}
\usepackage{flushend}
\usepackage{multicol}
\usepackage{enumitem}
\usepackage{supertabular}
\usepackage{hyperref}
\usepackage{hyphenat}
\usepackage{multirow}
\usepackage{url}
\usepackage{outlines}
\usepackage{multirow}
\usepackage{adjustbox}
\usepackage{listings}

\usepackage{xcolor}

\AtBeginDocument{%
  \providecommand\BibTeX{{%
    \normalfont B\kern-0.5em{\scshape i\kern-0.25em b}\kern-0.8em\TeX}}}

\begin{document}

\title{MIDMod-OSN: A Microscopic-level Information Diffusion Model for Online Social Networks}


\author{Abiola Osho}
\affiliation{%
 \institution{Kansas State University}
 \streetaddress{Rono-Hills}
 \city{Manhattan}
 \state{Kansas}}
\email{aaarise@ksu.edu} 

\author{Colin Goodman}
\affiliation{%
  \institution{Kansas State University}
 \streetaddress{Rono-Hills}
 \city{Manhattan}
 \state{Kansas}}
\email{c3544g@ksu.edu}

\author{George Amariucai}
\affiliation{%
  \institution{Kansas State University}
 \streetaddress{Rono-Hills}
 \city{Manhattan}
 \state{Kansas}}
\email{amariucai@ksu.edu}

\begin{abstract}
 As online social networks continue to be commonly used for the dissemination of information to the public, understanding the phenomena that govern information diffusion is crucial for many security and safety-related applications, such as maximizing information spread and misinformation containment during crises and natural disasters. In this study, we hypothesize that the features that contribute to information diffusion in online social networks are significantly influenced by the type of event being studied. We classify Twitter events as either informative or trending and then explore the node-to-node influence dynamics associated with information spread. We build a model based on Bayesian Logistic Regression for learning and prediction and Random Forests for feature selection. Experimental results from real-world data sets show that the proposed model outperforms state-of-the-art diffusion prediction models, achieving 93\% accuracy in informative events and 86\% in trending events. We observed that the models for informative and trending events differ significantly, both in the diffusion process and in the user features that govern the diffusion. Our findings show that followers play an important role in the diffusion process and it is possible to use the diffusion and OSN behavior of users for predicting the trending character of a message without having to count the number of reactions.

\end{abstract}


\settopmatter{printacmref=false}
\renewcommand\footnotetextcopyrightpermission[1]{} 
\pagestyle{plain} 

\maketitle

\section{Introduction}

Online social networks (OSN) have become increasingly important for the dissemination of information for public health, as well as during disasters and crises. While the dissemination of accurate information may protect the general public and potentially save lives, the spreading of false or inaccurate information is detrimental to public health and safety in those contexts. During a time when the spread of misinformation is an increasingly serious problem, it is important to study the creation and spread of information, as well as opinion formation in OSNs. To effectively establish this phenomenon, it is essential that we identify the key features that contribute to the repost and eventually, the spread of information in OSNs. Information diffusion describes how information is transmitted between individuals. In online social network like Twitter, with 335 million monthly active users as of the end of the second quarter of 2018 ~\cite{stat}, information can easily become viral because it allows strangers to filter, discuss and share information of common interest with networks of followers and through the use of hashtags. The ease of accessibility and the broad reach makes Twitter a strategic tool for businesses, interest groups, politicians and journalists and during crises and disasters.

Information is said to propagate, or diffuse, when it flows from one individual or community in a network to another. In the case of Twitter, diffusion can be seen as an action to share a Tweet with a user's followers with (i) no other new content added, called Retweet or (ii) new content added, called a Quote. Most studies in analyzing information diffusion focus on the overall spread of information by focusing on event detection and the spread of the event across the network without comprehensively evaluating the diffusion process on a microscopic level -- i.e., the factors that influence diffusion, differences in the spread of information in varying Twitter events and the information dissemination process. It is usually hard to assess why some information disseminates and other does not, but it is safe to assume that the features and/or the contexts of messages that go "viral" and those that do not must differ to some extent. In crises/trending Twitter events, the volumes of messages and interaction grow exponentially within a short time. This kind of interaction explosion is expected to impact the prediction model in a different way than when the spread is over a longer period. We assume that building a temporal pattern of a user's online behavior -- like the time of day when the user creates or reacts to tweets versus when the tweets get retweeted -- is important, as this behavior can be exploited for targeted information spread. By successfully identifying the features that make a difference in determining the virality status of a post, organizations can identify attributes to look out for in nodes that will ensure maximum information spread and nodes to avoid in case of containment. After predicting the spreading behavior of a post from one node to another, one can extend the scope of prediction to community-wide and/or network-wide.


Existing models for predicting information diffusion observe diffusion on a holistic level across trending events or hashtags. Many of these studies are focused on finding super-seeders, or influential nodes, based on the assumptions that the influence of the feature vector will be static across event types. The feature vector is a combination of attributes, possibly specific to user, message, network, and/or interaction, that contribute to an account's online persona. In this study, we hypothesize that the features that contribute to information diffusion in online social networks are significantly influenced by the type of event being studied. Since Twitter is increasingly becoming a place to visit for trends and breaking news, as well as asking questions and gathering information on general topics, we classify Twitter events as (1) informative for topics relating to general knowledge and which have not attained viral status, and (2) trending for topics that can be described as viral, breaking news, hot topics, or crises. We describe a topic to be trending if there are observed sharp spikes in the rate of posts relating to the topic instead of a gradual growth observed over a period of time. Similar to studies on predicting extremism ~\cite{ferrara2016predicting} and temporal dynamics ~\cite{guille} in social networks, we build a model that predicts diffusion using features learned from Twitter data. We go the extra mile by exploring the node-to-node influence dynamics associated with information spread. We use machine learning models to observe the performance and effect of similar sets of features on the previously identified Twitter events types to understand how the pattern of discussion, diversity in opinion, urgency and timeliness of topics influence diffusion behavior. The proposed model is built on Bayesian Logistic Regression for learning and prediction and Random Forests model for feature selection. These two statistical models have been observed to perform sufficiently well in predicting information spread in online social networks. The contributions of the paper are as follows:
\begin{itemize}
    \item We build a tool to crawl the Twitter Search API using user IDs and build a database encoded using JSON based on key-value pairs, with named attributes and associated values. We make replicability of our results possible by making the crawler and model publicly available. 
    \item We present a node-to-node feature analysis model to learn the diffusion process by combining a set of network, interaction, semantic and temporal features.
    \item We fit a stochastic model to the relationship between these features and the probability of diffusion.
    \item We identify the optimal subset of features needed to efficiently predict information diffusion in Twitter events.
    \item We draw conclusions regarding the best time to tweet, as well as the most important user attributes that contribute to achieving maximum retweetability, and in turn maximum diffusion, in the network. 
    \item We demonstrate the value of crowdsourcing to predicting the virality of a post before this would even be possible by counting user reactions.
\end{itemize}
The paper is organized as follows. Section \ref{prevwork} reviews the related work on information diffusion in social networks. Section \ref{model} describes our general approach, classification algorithms and evaluation metrics. Section \ref{experiment} describes the data collection, the experiment setup and the feature set, and presents our results. Section \ref{crowd} describes crowdsourcing on Twitter and how to predict if a post will go viral without information about the reaction count.Finally, Section \ref{conclusion} gives conclusions and insights into possible future works.

\section{Previous Work}\label{prevwork}
Methods for predicting information diffusion depend greatly on efficient topic and event detection, as well as feature selection. Within the vast literature on diffusion in networks relevant to our study, we provide a brief overview on information propagation and diffusion prediction models, with some details on recent work in feature selection for information diffusion.

\subsubsection{Information diffusion in social networks}
The information diffusion process can be observed through the diffusion graph and rate of adoption of the information by the nodes in the graph. The diffusion graph shows influence in the network, which is important for viral marketing \cite{subramani2003knowledge} \cite{chen2010scalable}\cite{domingos2005mining}, crisis communication \cite{acar2011twitter} and  retweetability\cite{neppalli2016retweetability}. Generally, influence analysis models have focused on relationship strength based on profile similarity and interaction activity \cite{xiang2010modeling}, and the mechanisms responsible for network homogeneity \cite{lewis2012social}. 

Identifying influential users has been found to be useful when trying to select seed nodes in the community that will maximize the spread of information across the networks. For instance, \cite{3} worked on finding the best spreaders in dissimilar social platforms when the complete global network structure is unavailable. The work of \cite{11} observed that (1) the authority of an  influential user on social media which can be used to change the opinions of other users and (2) opinion similarity factors where users tend to accept an opinion that is similar to his own, are important factors when selecting seed nodes for information spread. 

\subsubsection{Diffusion prediction models}
Predictive models like the independent cascade (IC) model \cite{kempe2003maximizing} make use of submodular functions to approximate the selection of most influential nodes. An inactive node $v$ can be activated by an active node $u$ independently with a probability $P_{[u,v]}$ at time $t$. People observe the choices of others and make decisions based on that observation while ignoring their personal knowledge. 
The linear threshold model (LT) deals with binary decisions where a node has distinctly mutually exclusive alternatives. At each discrete time period $r(t)$, an inactive node is activated by its already activated neighbors, if the sum of influence degrees exceeds its own threshold. A node's threshold at time $t+1$ is defined as the number of nodes who had made a decision at time $t$ \cite{granovetter1978threshold}.

Asynchronous IC and LT (AsIC and AsLT respectively) were defined in \cite{saito2010generative} and \cite{saito2008prediction}, by introducing a time delay parameter before a parent node can activate an inactive child node. In AsIC, if the child node remains inactive after the specified period $\delta$, the parent node is given only a single chance to attempt activating the child node to eliminate the likelihood of a single node being simultaneously activated by multiple parent nodes. In AsLT, a node decides when to receive the information once the activation condition has been satisfied. The diffusion process unfolds in continuous-time $t$, and proceeds from a given initial active set such that if the total weight from active parent nodes exceeds $\theta_v$ at time $t$ for the first time, $v$ becomes active at time $t + \delta$.

Some other studies like \cite{wang2012diffusive} propose a model based on Partial Differential Equations (PDE) by introducing a diffusive logistic to model to predict temporal and spatial patterns in Digg, an online social news aggregation site. A Linear Influence Model was developed in \cite{yang2010modeling},  focused on modeling the global influence of a node on the rate of diffusion through the implicit network by estimating an influence function to quantify the number of successive adoptions attributed to a node over time.

\subsubsection{Feature selection for information diffusion}
\cite{guille} introduced a variant of the AsIC model called the T-BAsIC framework that assigns a fixed value for a real time-dependent function for each link, without fixing the diffusion probability. The model relies on three different dimensions to compute the diffusion probability: social, semantic, and time. The model was designed to predict the daily volume of tweets for a topic and variations in popularity of topics over time. They proceeded by identifying 2 types of users: (1) transmitters that pass along information and (2) stiflers that become dead-ends for information travel, with stiflers growing with time for a given topic. 

In \cite{ferrara2016predicting}, the authors leverage a mixture of metadata, network and temporal features in detecting users spreading extremist ideology and predict content adopters and interaction reciprocity in social media. They adopted logistic regression and random Forests learning models with 52 features observed from Twitter data of over 25,000 accounts labeled as supportive of the Islamic State. 
Given the temporal relevance of tweets, \cite{spasojevic2015post} propose  finding the best times for a user to post on social networks in order to maximize the probability of audience response. They hypothesize that the probability that an audience member reacts to a message depends on factors such as his daily and weekly behavior patterns, his location and timezone, and the volume of other messages competing for his attention.

Instead of focusing on the diffusion of trending events, our model seeks to show the difference between the diffusion models of informative and trending Twitter events, as it relates to the volume of posts, features influencing diffusion and time to post for maximum spread. Also, we enhance the model to predict the virality of a post by adopting crowdsourcing techniques.

\begin{table*}[ht!]
\centering
\begin{tabular}{lllllc}  
\toprule
Event type & Topic  & No. of Users & No. of edges & No. of tweets & diffused/not diffused ratio\\
\midrule
\multirow{2}{*}{Informative} &  Health benefits of coffee    &  50919 &  1100270  & 2958382 &\multirow{2}{*}{40/60}\\
            &   Mental health &  29362 &  3224330  &  4030412  \\
\midrule
\multirow{2}{*}{Trending}   &   2018 Kansas general elections &  15339  & 2509255  &  24188962 & \multirow{2}{*}{52/48}\\
            &   Government shutdown  & 12581 & 2549136  & 14513377\\
\bottomrule
\end{tabular}
\scriptsize\caption{Data distribution}
\vspace{-5mm}
\label{tab:data}
\end{table*}


\section{Model and method}\label{model}
In this section, we describe the dataset and data gathering process, learning and feature selection algorithms, as well as the evaluation metrics for our microscopic-level information diffusion model for online social networks (\emph{MIDMod-OSN}).

\subsection{Dataset description} \label{dataset}
One of the biggest challenges to this research is access to data, as most of the datasets and tools available only provide part of the information (usually, tweets and network features) needed for academic research. Due to the number of features being examined, we needed the complete metadata of Tweet and user JSON  (JavaScript Object Notation) objects. For the purpose of future research requiring Twitter JSON objects, we created a tool that  crawls the Twitter Search API using the usernames or IDs of a set of seed users and made it publicly available on GitHub. The tool creates a relationship graph built around the seed users and their followers. Since it is almost impossible to have the complete Twitter graph, the sub-graph generated is as representative as it can be. For each of the 4 topics we are exploring, we randomly select 50 users and build a followership relationship around them for up to depth 2. The user (or node, used interchangeably throughout the remainder of this paper) information is then used to build a database crawled over a 30-day period, by collecting all the tweets created by users in the sub-graph during this time period. 

The use of Twitter to report real-life events is steadily increasing and for this study, we classify these events into two categories: informative and trending. We then base our study on two different topics for each event. The topics defined are (1) Informative: (1.1) Health benefit of coffee, (1.2) Mental health and (2) Trending: (2.1) 2018 Kansas elections, (2.2) Government shutdown. The data and network distribution for the dataset can be found in Table~\ref{tab:data}. We associate each topic with a bag of words that are deemed important to the topic by creating a list of words frequently used with or associated with the topic. A tweet is said to be relevant to a topic if and only if it contains one or more of the predefined keywords. For example, 60 key words were used to identify tweets belonging to the topic of \emph{health benefit of coffee}. The data is split into (1) a training set used for parameter estimation and (2) a test set to assess the performance of the model.

\noindent
\textbf{Information spread behavior:} In a directed network $G = (V, E)$ with no self-links (communities within the graph might contain cycles), $V$ is the set of nodes and $E (\subset V \times V$) is the set of edges. For each node $v \in V$, we denote $U$ as the set of $v$'s followers and $W$ as the set of $v$'s friends, \textit{i.e.,} $U = \{u; (v, u) \in E \}$ and $W = \{w; (w, v) \in E \}$, respectively. Similar to \cite{saito2010generative}, we assume \textit{AsIC} with the time delay function associated with information diffusion along the edge. At time $t$ each node $v$ gets a chance to activate (get a reaction through retweet, favourite, quote or reply) its follower $u$. If  node $u$ is not activated by time $t + \delta$, then node $v$ looses the competition for activating $u$ to any other node $v'$ that attempts to activate $u$ between $t+\delta$ and the time of $u$'s activation. For simplicity, we assume that activation is restricted to a node's interaction with the network, but in reality, this will not always be the case, as activation is not solely dependent on the network activities but could be from sources external to the network itself, thereby causing delay in activation.

\subsection{Learning and feature estimation models }
The model we propose takes a pair of users with established followership relationship and extracts a set of attributes classified as: Network, Interaction, Semantic and Temporal. We adopt two off-the-shelf machine learning models: Bayesian Logistic Regression and Random Forests due to the good performance of both models in similar settings, as observed in \cite{guille} and \cite{ferrara2016predicting}. First, we use the attributes described in Table ~\ref{tab:features described} to train our model based on Bayesian Logistic Regression (BLR). The prediction capability of the model is tested and evaluated before the feature selection phase. One challenge with high dimensional models is that as dimensionality increases, the space between data points become very large \cite{beyer1999nearest}, to the extent that it is difficult to produce reliable results. By removing features that are highly correlated and those with minimal effect on the predictability of the model, we select a subset of the original features by using Random Forests (RF) as a filter. The BLR model is then re-trained with the selected feature set and evaluated to determine the predictive abilities of the selected features.

\begin{figure}
    \includegraphics[width=\linewidth]{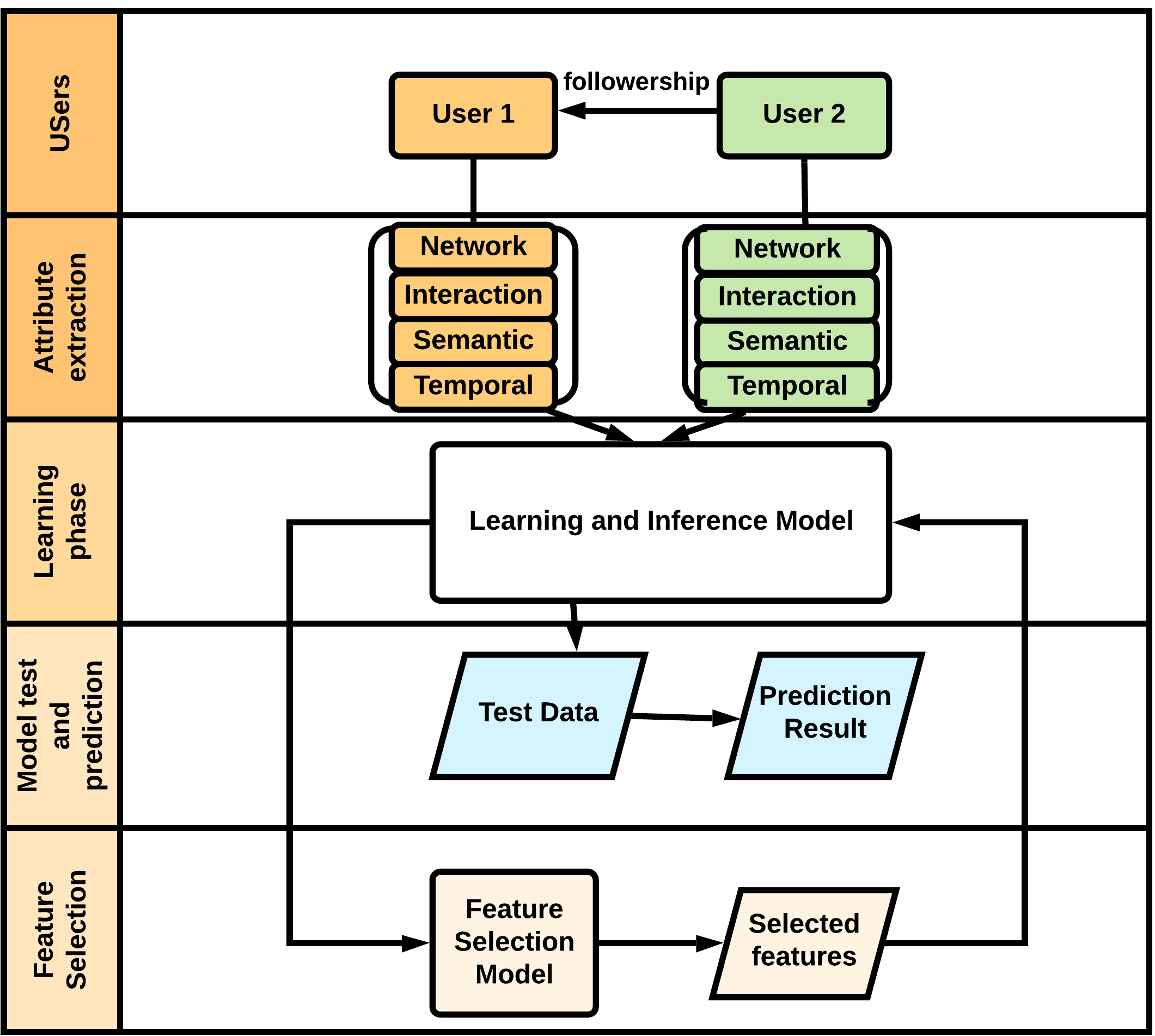}
    \caption{Illustration of MIDMod-OSN's approach to predicting user 2's reaction to user 1's post}
    \vspace{-5mm}
    \label{fig:methodology}
  \end{figure}

We perform node-to-node influence analysis by examining feature performance between two users with established followership relationship. We extract attributes from our dataset and organize them as: Network, Interaction, Semantic and Temporal, see Figure \ref{fig:methodology}. The features are estimated for both the source and destination nodes, with an associated binary label depicting diffusion along the edge between them. For each user, we learn 27 features, and a social homogeneity (common to two users, showing an overlap in the sets of users they relate with, i.e. common friends and followers) by adopting the features of \cite{guille} (excluding the temporal feature) and introducing new ones. Since each observation is a pair of users given as source and destination, the input to the learned model is a vector of 55 features along with a diffusion label per data point. For the temporal dimension, we study the creation, consumption and forwarding of content by splitting a 24-hour period into 6 hours interval (i.e., 0:00-5:59 am, 6:00-11:59 am, 12:00-5:59 pm, 6:00-11:59 pm) and learn a model for each time period. Overall, we learn 4 temporal models for each pair of users, to observe how the post and reactions to post behaviors change across different time periods in a day.

\begin{table*}[ht]
\begin{tabular}{p{0.5in}p{2.1in}p{3.4in}}  
\toprule
Feature class & Feature  &  Description \\
\midrule
\multirow{3}{*}{Network} & avg number of followers & higher follower count depict higher reach\\
    & avg number of friends & average number a user follows\\
    & ratio of followers-to-friends& shows how balanced the user's network is \\
\midrule
\multirow{18}{*}{Interaction} & volume of tweets & normalized over account's lifetime\\
     & social homogeneity & shows two users' common friends and followers\\
     & ratio of directed tweets & percentage of his posts are directed at others\\
     & active interaction & binary value depicting established interaction between them\\
     & mention rate & gives volume of posts directed at the user\\
     & ratio of retweet-to-tweet & percentage of user's posts that have been retweeted\\
     & tweets with hashtags & how many of his original posts contain hashtags\\
     & retweets with hashtags & shows the user follows and reacts to posts containing hashtags\\
     & volume of retweets over account's lifetime & we assume the account is a forwarding bot if all his posts are retweets\\
     & avg tweets per day & gives insight into how active the user is\\
     & avg number of mentions excluding retweets & shows how interesting others find the user\\
     & ratio of mentions-to-tweet & includes posts where the user mentioned and retweeted other people's posts\\
     & tweets containing URL & shows how many of the user's original tweet contain URLs\\
     & retweets containing URL & shows the user follows and reacts to posts containing URLs\\
     & tweets containing media & shows how many of the user's original tweet contain media (photos, videos)\\
     & retweets containing media & shows the user follows and reacts to posts containing media\\
     & presence of user description & a boolean value showing if the user's profile has description (bio)\\
     & ratio of favorited-to-tweet & shows how many of the user's tweets have been endorsed by others\\
\midrule
\multirow{3}{*}{Semantic} & presence of keywords & boolean value that shows if the user has tweeted about the topic\\
     & positive polarity of tweets & percentage obtained from running sentiment analysis on all the user's tweets\\
     & negative polarity of tweets & percentage obtained from running sentiment analysis on all the user's tweets\\
\midrule
\multirow{4}{*}{Temporal} & ratio of tweets per time & percentage of all user's posts within a time period\\
     &ratio of tweets that got retweeted & percentage of original tweets that got retweeted within a time period\\
     & ratio of retweet per time period & percentage of reactions user produce within a time period \\
     & average time before retweet & estimates the average time elapsed before the user gets a reactions\\
\bottomrule
\end{tabular}
\caption{Features extracted for each user to serve as input variables to the learning model}
\vspace{-5mm}
\label{tab:features described}
\end{table*}

\subsection{Model Evaluation} \label{eval}
Each input is a vector set of 55 features, learned over 4 different time periods. The performance of the models are obtained using the $k$-fold cross validation technique, with $k = 10$ folds, and using the $80\% - 20\%$ training-test data split and averaging performance across the 10 folds.
The prediction capabilities of the learned model are tested based on its abilities to predict if there is diffusion across an edge given the learned model. We employ standard machine learning evaluation metrics: Precision, Recall and F1 score, along with Area under the Receiver Operating Characteristics (ROC) curve to measure the predictability of the model. 



\section{The Diffusion Prediction Experiment}\label{experiment}
In this section, we describe our experimental setup, and the results obtained for each phase of our model. We evaluate the performance of the prediction and feature selection models, and then make comparisons with state-of-the-art prediction models. Finally, we discuss the time to tweet paradigm based on our observations.

\subsection{Experimental setup}
We perform a supervised learning task where we train the model using the attributes from a pair of nodes with an established followership relationship and label the interaction between them as either diffused or not diffused. An edge is said to be diffused if and only if the destination user (in Twitter terms: \textit{follower}) has at any point forwarded his friend's (\textit{followee}) messages on the topic being examined. The attributes learned are said to be representative of users' network, interaction,  participation, role and importance in the spread of information to other nodes in the network. As previously stated, these attributes are learned over four different time intervals. After learning these features, we fit a regression function that maps the learned user attributes to the likelihood of diffusion between the nodes. 

Given the directed nature of the Twitter graph, the learning task is non-deterministic, as switching the source and destination nodes may produce a different mapping between the input and output variables. Initially, we maintain an equally weighted feature space with the assumptions that each feature will influence the forwarding decision (reshare, reply or not) with equal magnitude. Subsequently, the feature selection framework is initialized to first learn a function with the same set of attributes, secondly rank the features in decreasing order of importance, and third retrain the model using the 15 most important features. 


We evaluate the effectiveness of our model and methods on predicting diffusion between node pairs in the spread of information across the social network on selected topics using the methods described in Section \ref{eval}. Also, we present our findings on the optimal subset of features necessary for maximized diffusion predictions, with discussions on the best time to post given the event type. Experimental result show a significant improvement over state-of-the-art models both in accuracy of prediction and the ability of the model to differentiate between diffused and not diffused edges.

\subsection{Diffusion prediction model} \label{diff prediction}
Firstly, we observed that the volume of tweets across a 30-day period varied widely for informative and trending events. As shown in Table~\ref{tab:data}, it can be established that even though the combined number of users observed in the trending events is 2.8 times less than the number of users across informative events, we were still able to record 5.5 times more tweets over informative events. We note that in our dataset, trending events generate up to 15 times more tweets than informative events with the same network size. This sort of data projection will be sufficiently affected by the impact of the topic. For instance, one can forecast such data growth for trending events with wide reach like political and health topics but not in lifestyle. Other factors that will impact the data projection include time of day, and external sources like coverage in traditional media.

In Table~\ref{tab:precision}, we show the performance of our models, averaged out across topics in each event class, given the performance metrics previously highlighted. Using the F1 measure, the model achieved 93\% accuracy in prediction in informative events and 86\% in trending events. The simplified models, based on the 15 most important feature for training, showed a 90\% prediction accuracy in informative events and 89\% in trending events. Results in the present study are consistent with the prediction results for trending events in past literature.

\begin{table}[ht!]
\centering
\begin{tabular}{llccc}  
\toprule
Event type & Model  & Precision & Recall &  F1  \\
\midrule
\multirow{2}{*}{Informative} &  55 features   & 0.91  &  0.96  & 0.93 \\
            &  top-15 & 0.87  &  0.94  &   \textbf{0.91} \\
\midrule
\multirow{2}{*}{Trending} &  55 features   & 0.87  & 0.84   &  0.86\\
            &  top-15 & 0.89  & 0.90    &  \textbf{0.89}  \\
\bottomrule
\end{tabular}
\scriptsize\caption{Performance evaluation of MIDMod-OSN in predicting diffusion of posts from different event types.}
\vspace{-4mm}
\label{tab:precision}
\end{table}

Furthermore, we compare both our prediction models with the state-of-the-art diffusion prediction model proposed by Guille et. al. \cite{guille}, see Table~\ref{tab:comparison}, and observe that both models with 55 and top-15 features, perform considerably better than the state of the art. Our hypothesis that increasing the feature vector space by extracting and learning more attributes from the Twitter JSON objects will make the predictive model more robust is proved correct as we were able to record a 7\% increase from the model of Guille et al. It might be argued that a 7\% increase is not enough to justify the increase in computation time and resources caused by the increase in feature space, however, we oppose this argument with the feature selection phase, introduced solely for maximizing diffusion prediction by utilizing the features that will directly impact the information spread. For a small cost in accuracy, reducing the input variables by 72\% (top-15 features) will give a prediction accuracy of 91\%, which is only a 2\% reduction in predictive power (when compared with all 55 features). In like manner, an 81\% reduction (top-10 features) yields a prediction accuracy of 87\%, constituting a 6\% reduction in accuracy. The trade-off in adopting the top-10 features is significant, and as such, we adopt the top-15 important features as the optimal set of features necessary for diffusion prediction without incurring expensive computational costs. 

\begin{table}[ht]
\centering
\begin{tabular}{llcc}  
\toprule
Event type & Model  & F1 & AUC-ROC \\
\midrule
\multirow{4}{*}{Informative} &  55 features   & 0.93  &  0.98   \\
            &  top-15 & 0.91  &  0.96  \\
            & top-10 & 0.87 & 0.94 \\
            & Guille et al.(13 feat.) & 0.86 &0.94 \\
\midrule
\multirow{4}{*}{Trending} &  55 features   &  0.86 &   0.94  \\
            &  top-15 & 0.89 &   0.96  \\
            & top-10 & 0.88 & 0.94 \\
              & Guille et al.(13 feat.) & 0.88& 0.95\\
\bottomrule
\end{tabular}
\caption{Prediction accuracy using proposed model with different number of features and state-of-the-art.}
\vspace{-3mm}
\label{tab:comparison}
\end{table}

Contrary to expectations, it is observed that learning all possible features in trending events impacts prediction accuracy negatively. Due to the consistently changing pattern of interactions and behavior in trending events, increasing the number of features learned brings about over-fitting caused by the exponential growth in the data needed for training. We are able to mitigate the impact of over-fitting in the model using the $k$-fold cross validation technique, with $k$ set to 10. Nonetheless, it will be detrimental to suggest that learning these features is of no value, as we are convinced that feature selection over several topics will be useful in building a template of attributes for a pre-trained prediction model. The accuracy of the prediction model is consistent with previous studies that have focused on Trending events.

\begin{table*}[ht!]
\centering
\begin{tabular}{c l l} 
\hline
Rank & Informative  & Trending\\
\hline
1 & dest (destination node) average url per tweet    &  Social homogeneity    \\ 
2 &src (source node) ratio of retweet per time period & dest active interaction between the nodes      \\  
3 & src volume of tweets over account’s lifetime  &    src  avg number followers    \\                
4 & dest ratio of tweets that got retweeted per time period    &  src ratio of favorited to tweet\\
5 & social homegenity             &   src mention rate  \\        
6 & dest avg number of media in retweets    & src ratio of retweet per time period\\              
7 & src ratio of retweets to tweets &   src volume of tweets over account's lifetime    \\  
8 & src ratio of tweet per time      & src active interaction between the nodes              \\        
9 & src ratio of tweets that got retweeted per time period     &  src avg url per tweet\\
10 & dest avg number of retweets with hastags      &     src ratio of retweets to tweets   \\          
11 & dest ratio of retweet per time   &     src ratio of mentions to tweet\\               
12 & src avg number of retweets with hastags   &    src avg number of tweets \\               
13 & src average url per retweet              &    src avg number of mentions not including retweets  \\                
14 & src avg number of  tweets         &     dest avg number of mentions not including retweets  \\                    
15 & dest avg number of retweets  & dest volume of tweets over account's lifetime  \\
\hline
\end{tabular}
\scriptsize\caption{Ranking of the top 15 optimal features that should be maximized for maximum diffusion or minimized for containment}
\vspace{-3mm}
\label{feat_select}
\end{table*}



\subsubsection{Cross testing between models}
To further show that the performance of the models is not biased to topic domains, we tested the informative model with a political related topic and trending model with an health related topic. On testing both models with data from new topics (not used for training and in new topic domains), we observed results similar to those reported earlier with F1 score of 90.1\% for informative and 89\% for trending events. This confirms that the models will perform comparably regardless of topic domain. 

To ascertain that there is indeed a difference between the informative and trending models, we evaluated the informative model with data from trending topics and evaluated the trending model with data from informative topics. The objective is to test if the knowledge gained from one model can be used in making predictions in the other. The outcome of predicting the diffusion of trending posts using a trained informative model produced an F1 score of 82\%, while we observed an F1 score of 78\% from predicting informative posts using a trained trending model. This result is not totally surprising, due to the irregular pattern associated with posts and users contributing to trending topics. 


\subsection{Feature selection framework}
One justification for using multivariate methods is that they take into account feature redundancy and yield more compact subsets of features, as features that are individually irrelevant may become relevant when used in combination, which also shows that correlation between sets of features does not necessarily imply redundancy. 
Evaluating the Random Forests model using a 10 fold cross-validation technique achieved an AUC score of 99\%  in both informative and trending events using the complete set of features. Considering that the goal of the feature analysis task of this study is to identify the optimal set of features necessary to maximize diffusion prediction, we select the top 15 features, rather than the traditional top 10 (for reasons highlighted in \ref{diff prediction}). In Table~\ref{feat_select}, we report the ranking of the top 15 features in the two event types. 


Given two users, we observed that the attributes of the followers (destination nodes) account for 40\% of the optimal subset of features, in informative events, and for 20\% in trending events. In recent happenings in online social networks, it has been observed that discussions and threads that impact trending events are not usually trending in nature. For instance, the much publicized propaganda campaign during the U.S. 2016 elections targeted users on both sides of the political divide by exposing them to opinions formulated over time, using hashtags and shortened URLs. In real life, a considerable number of trending topics are indeed informative events that become trending due to a change triggered by an incident. Irrespective of the type of event, social homogeneity and source's (1) ratio of retweet per time period (2) volume of tweets over account’s lifetime (3) ratio of retweets to tweets (4) average number of tweets, prove to be important in the information diffusion process. 

We notice that the follower's features are powerful enough to impede diffusion in informative events but these abilities diminish as the event becomes trending. As topics become viral, the number of followers a user has ranks third in trending events. Even though this feature is previously deemed unimportant in informative posts, combining it with a high ratio of retweets to tweets, mention rate and active interaction from his follower will boost his reach. It is inadequate to assign importance to an account across all networks and topics, as seen in \cite{rao2015klout}, if the importance and authority it wields vary with changing topic, event and social network. It is paramount that the relevance of a user be decoupled across social networks, especially Twitter, since a considerable number of users maintain a level of anonymity. For instance, a user will not run a web search on an account to confirm the authenticity or authority of its posts before reacting on Twitter. Also, a user that is authoritative on  health-related issues on Twitter might be an unreliable source of health-related posts on Facebook. It is inadequate to assign importance to an account across all networks and topics if the importance and authority it wields vary with changing topic, event and social network. Throughout this research, we demonstrated that the role of the followers in diffusion prediction is more than just a contribution to the follower count of the sender, and should combine the effectiveness of the interaction of each follower with their friend. Our results show that the influence a user wields in a network is an aggregate of his influence over each of the nodes in the network, thus combining all three centrality concepts as introduced by \cite{freeman1978centrality}.

\subsection{Time to tweet}
The results from our experiment validate our assumptions that the extent to which messages diffuse will be significantly influenced by the time when they are created. As observed in the top-15 most important features, see Table ~\ref{feat_select}, for both the follower and followee in the network, the time period where most of their messages (original tweets and reaction) fall are crucial to propagation. Experimental results show that more than 75\% of informative posts fall into the 2nd (6:00-11:59 am) and 4th (6:00-11:59 pm) time periods, but those of trending posts are in the 3rd (12:00-5:59 pm) and 4th (6:00-11:59 pm) time periods. It is interesting to note that both Twitter event types got considerable attention during the 4th time period as this for most people is a time to catch up with the day's activities.
However, we observed that the best time to tweet an informative message on Twitter for maximum diffusion is in the 2nd time period, while trending is in the 4th. We speculate that the contrast in peak diffusion times can be due to the reactive nature of trending events, occurring mostly after the day's activities, unlike the active nature of informative events, where a user is mostly putting opinion out.

\section{Crowdsourcing for Early Trending Topic Detection} \label{crowd}
In this section we discuss the concept of crowdsourcing in OSNs, and why it is important. We describe the experiment and experimental results on adopting MIDMod-OSN for crowdsourcing the early detection of trending topics.

\subsection{The early detection of trending topics}
Individuals and organizations looking to use Twitter as an advertising or political campaign platform will find it useful to know ahead of time if a newly created message or hashtag will become trending, in order for them to maximize the attention for personal gain or minimize negative exposure. Similarly, governmental or non-governmental organizations attempting to neutralize the spread of misinformation during crisis scenarios could monitor users' reactions to previously identified harmful-misinformation-carrying messages, and predict whether these messages will become viral before this determination can be done via standard methods, like counting tweets. This would enable them to effectively fight the further spread of the misinformation before it has a chance of becoming viral.

\subsection{Using MIDMod-OSN for crowdsourcing}
In the past, individuals and organizations have used OSNs like Twitter as an avenue to obtain ideas in a crowdsourcing context. In this study, we view a user's reaction to a post as an implicit contribution towards crowdsourcing, where users' posts and reactions to posts serve as a form of criticism or validation, report on crisis and event, advert of product and service, protest, or even political campaign.

Users react uniquely to posts, and their reaction may or may not be correlated with the message's potential for becoming trending. While some users react to posts from all event types (trending and informative), others only react (share, quote, favorite, reply or retweet) to tweets that are trending or about to attain the trending status because of the need to share or contribute to hot topics. This kind of users can serve as discriminants in the model that predicts the trending character of a message. The goal of the prediction task is to show that the diffusion behavior and OSN behavior of users is useful for predicting the trending character of a message when the reaction count is unavailable.

\subsection{Experiment design and results}
For this experiment, we are interested particularly in evaluating the usefulness of users' reactions to predicting message virality. It is for this reason that we must avoid (1) including specific message features in the classifier and (2) including -- explicitly or implicitly -- counts of tweets relating to a specific message. The first requirement fits naturally with our previous model, which only relies on user, rather than message, features. To satisfy the second requirement, we must construct an experiment that treats each user interaction with the message independently of all others. That is, we purposely make a prediction of virality from each user interaction, rather than combining all user interactions into a single model.

Our model predicts if a message will go viral or not, by including the diffusion property \emph{diffuse/not diffuse} of the message as an independent variable during the training phase. We examine how users on Twitter relate with posts of their friends by building a classifier to distinguish user interactions based on the virality status of the message. For a message $m$, where $m \in \{1,\ldots,M\}$, spread over a network with $n$ interactions, we train a model that predicts the virality status of the message based on the diffusion behavior observed along each one of the $n$ links along which the message propagates. This results in $n$ distinct predictions. The overall predicted output is calculated as the majority virality status observed across the $n$ interactions. We select 1000 messages --500 each-- from trending and informative event types and evaluate the MIDMod-OSN's ability to predict if a message will go viral or not. For instance, if a trending message is spread over 5 interactions and the model predicts the post to be Trending 3 out of 5 times, we accept the output as Trending and evaluate the model over its correct classification of $M$ messages in the test collection. 

We run the experiment with 10000 users. With this fraction of the network, we were able to show that to a certain degree that the diffusion behavior and OSN behavior of users is useful for predicting the trending character of a message without having to count the number of reactions, see Table \ref{tab:crowdsource}. 

\begin{table}[ht!]
\centering
\begin{tabular}{lccc}  
\toprule
Model & Precision & Recall &  F1  \\
\midrule
virality-predicting & 65 & 78 & 70 \\
\bottomrule
\end{tabular}
\scriptsize\caption{Performance evaluation of MIDMod-OSN in predicting the trending status of a message without counting reactions.}
\vspace{-4mm}
\label{tab:crowdsource}
\end{table}

We should note here that when attempting to predict message virality, one should consider a more comprehensive model, including message attributes and a joint treatment of all user reactions to a specific message. Nevertheless, the results of this experiment demonstrate that crowdsourcing (at least part of) the detection mechanism is not without merit.


\section{Conclusion and Future Work}\label{conclusion}
Predicting information diffusion will continue to be an active research area due to the fast growing importance of online social networks. In this paper, we studied the problem of identifying features that impact diffusion in different types of Twitter events. We created the MIDMod-OSN model and trained it using 55 features extracted directly from the Twitter REST API and outperformed the prediction power of state-of-the-art models. We further established that a prediction model based on the top-15 most important features, selected by our feature selection framework, is optimal in correctly predicting diffusion, achieving an AUC score of 96\% in both event types.
Our main theoretical contribution is  distinguishing between Informative and Trending Twitter events, and teasing out differences in information diffusion patterns. 
Even though they are generally overlooked, informative posts make up a big chunk of messages shared on social networks.
We showed the differences between the pattern of interaction between users when exchanging these kinds of posts and trending posts. Additionally, we establish the divergence in features influencing reaction to post, with 40\% of the top ranked features belonging to the followers in informative events and 20\% in trending events. From our results, we infer that for an influence maximization model to be effective, it needs to combine centrality concepts for control, efficiency and activity.

Future work may include more complex prediction tasks, involving the use of latent user and message attributes for predicting user reactions to posts based on the user's perceived veracity of the post in OSNs.

\bibliographystyle{ACM-Reference-Format}
\balance
\bibliography{midmodosn}

\end{document}